\documentclass[journal]{IEEEtran}
\usepackage{mathrsfs}
\usepackage{bbm}
\usepackage{amssymb}
\usepackage{bbding}
\usepackage{threeparttable}
\usepackage{diagbox}

\usepackage[table]{xcolor}
\definecolor{headerbg}{HTML}{DDEEFF}   
\definecolor{headertxt}{HTML}{003366}  
\usepackage{tabularx}

\usepackage[mathcal]{euscript}

\usepackage{psfrag,calc,url,bm}

\usepackage{cite}

\usepackage{graphicx}

\usepackage{psfrag}

\usepackage{subfigure}
\usepackage{hyperref}
\usepackage{url}

\usepackage{stfloats}

\usepackage{amsmath}

\usepackage{float}

\usepackage{algorithmic}

\usepackage[ruled,linesnumbered,vlined]{algorithm2e}

\usepackage{color}

\usepackage{boxedminipage}

\usepackage{amsthm}

\usepackage{multirow}

\usepackage{setspace}

\usepackage{soul}
\usepackage{epstopdf}

\allowdisplaybreaks[3]

\begin{document}

\title{Mamba for Wireless Communications and Networking: Principles and Opportunities }
\author{Rongsheng Zhang, Ruichen Zhang,~\IEEEmembership{Member,~IEEE}, Yang Lu,~\IEEEmembership{Member,~IEEE}, \\ Wei Chen,~\IEEEmembership{Senior Member,~IEEE}, Bo Ai,~\IEEEmembership{Fellow,~IEEE}, and Dusit Niyato,~\IEEEmembership{Fellow,~IEEE} 
\thanks{Rongsheng Zhang and Yang Lu are with the State Key Laboratory of Advanced Rail Autonomous Operation, and also with the School of Computer Science and Technology, Beijing Jiaotong University, Beijing 100044, China (e-mail: 24120413@bjtu.edu.cn, yanglu@bjtu.edu.cn).}
\thanks{Ruichen Zhang and Dusit Niyato are with the College of Computing and Data Science, Nanyang Technological University, Singapore 639798 (e-mail: ruichen.zhang@ntu.edu.sg,dniyato@ntu.edu.sg).}
\thanks{Wei Chen and Bo Ai are with the School of Electronics and Information Engineering, Beijing Jiaotong University, Beijing 100044, China (e-mail: weich@bjtu.edu.cn,boai@bjtu.edu.cn).}
}
\maketitle

\begin{abstract}
Mamba has emerged as a powerful model for efficiently addressing tasks involving temporal and spatial data. Regarding the escalating heterogeneity and dynamics in wireless networks, Mamba holds the potential to revolutionize wireless communication and networking designs by balancing the trade-off between computational efficiency and effectiveness.  This article presents a comprehensive overview of Mamba' applications in wireless systems. Specifically, we first analyze the potentials of Mamba for wireless signal processing tasks from the perspectives of  long-range dependency modeling and spatial feature extraction. Then we propose two application frameworks for Mamba in wireless communications, i.e., replacement of traditional algorithms, and enabler of novel paradigms. Guided by the two frameworks, we conduct case studies on intelligent resource allocation and joint source and channel decoding to demonstrate  Mamba's improvements in both feature enhancement and computational efficiency. Finally, we  highlight critical challenges and outline potential research directions for Mamba in wireless communications and networking. 
\end{abstract}


\section{Introduction}




%

Mamba  has emerged as a structured sequence modeling architecture\cite{SSM}, and achieved superior performance over Transformer-based models in various long-sequence applications, such as language modeling, audio generation, and DNA modeling\cite{Mamba}. Particularly, Mamba offers the ability to capture long-range dependencies with linear time complexity. Unlike previous deep learning (DL) models that treat spatial and temporal features in isolation, Mamba unifies these dimensions within its state space model (SSM), enabling seamless integration of sequential dynamics and structural dependencies. Thus, Mamba is  well-suited for both temporal tracking and spatial reasoning tasks. On the other hands, the enhancement of  intelligence in wireless communication systems has become an irreversible trend. Multi-functional wireless services and ultra-dense network topologies introduce complex heterogeneity and dynamics into wireless networks, and pose significant challenges to traditional signal processing approaches. Developing powerful DL approaches for wireless communications and networking  relies on the  match between wireless tasks and DL models. In light of the time- and space-varying characteristics of network topologies and the critical demand for real-time inference, there exists an inherent alignment between intelligent wireless designs and Mamba. Thus, investigating Mamba's applications in wireless communications and networking meets the vision of 6G.


Recently, a surge of DL approaches has emerged to tackle the escalating demands of wireless tasks, spanning channel estimation, semantic communication, signal classification, and resource allocation \cite{ResourceAllocation}. Diverse neural networks (NNs) have been deployed across wireless scenarios and tasks, demonstrating remarkable performance gains. For instance, convolutional NNs (CNNs)  achieved accurate channel estimation in ultra-massive multi-input-multi-output (MIMO) systems in \cite{CE-CNN}; recurrent NNs (RNNs) demonstrated strong performance in sequential tasks such as traffic prediction and symbol detection \cite{RNN}; graph NNs (GNNs) effectively modeled spatial dependencies in ultra-dense networks for scalable resource allocation \cite{GNN}. To pursue more powerful models, some studies also explored hybrid architectures combining these models to exploit complementary strengths. For example, the framework in \cite{DeepSIG} integrated CNN, RNN, and GNN within a unified architecture to handle heterogeneous inputs, while Transformer-based models such as NMformer \cite{NMformer} and DeepSC \cite{DeepSC} exhibit powerful sequence modeling for environmental signal classification and semantic communications, respectively. However, existing DL models often struggle to efficient handle spatial and temporal dynamics simultaneously. Mamba offers a unified and efficient architecture for jointly modeling spatial and temporal features. By leveraging a selective SSM, Mamba reduces temporal modeling costs with sequence length while preserving long-range dependency tracking. Its linear-time operations and continuous state transitions also enable efficient modeling of dynamic relationships across long sequences. For clarity, we summarize the comparative details in Table \ref{Table}. With all these potentials, Mamba is poised to revolutionize wireless network design by balancing computational efficiency and effectiveness.

Existing literature has comprehensively overviewed DL-enabled wireless intelligence. However,  the application of Mamba in wireless networks remains underexplored. To fill this gap, this article systematically analyzes the suitability of Mamba for wireless signal processing, followed by two proposed novel application paradigms. The contributions of this work is summarized as follows.
\begin{enumerate}

\item We highlight the key processes of Mamba, encompassing  dynamic state-space transition control, selection mechanism, and hardware-aware algorithm, and analyzes Mamba's advantages for long-range dependency modeling and spatial feature extracting for wireless communications.  

\item We propose two core application frameworks for Mamba in wireless communications and networking, i.e., replacing traditional algorithms to substitute high-complexity modules, and enabling novel paradigms to empower innovative processing frameworks for wireless communications. 

\item We present  case studies following the two application frameworks, i.e, Mamba-enabled resource allocation and Mamba-enabled joint source-channel decoding (JSCD). Experimental results demonstrate that the former significantly boosts inference efficiency with increasing  network scales, while the latter enhances semantic performance and maintain computational efficiency across various channel conditions.


\end{enumerate}

The rest of the article is organized as follows. We first introduce the technical principles of the Mamba architecture and its application in sequence and non-sequence modeling tasks. Then, we explore how Mamba can be applied to representative wireless communication problems, including resource allocation and JSCD. We present case studies to evaluate its efficiency and performance. Finally, we summarize the contributions of this paper and discuss potential future application directions of Mamba in wireless communication systems.


\section{Technical principle of Mamba}


\begin{table*}[t]
\renewcommand{\tablename}{Tab}
\centering
\caption{Multidimensional comparison of Mamba, RNN, Transformer, LSTM, and Hyena \cite{Hyena}.}
\includegraphics[width=\linewidth]{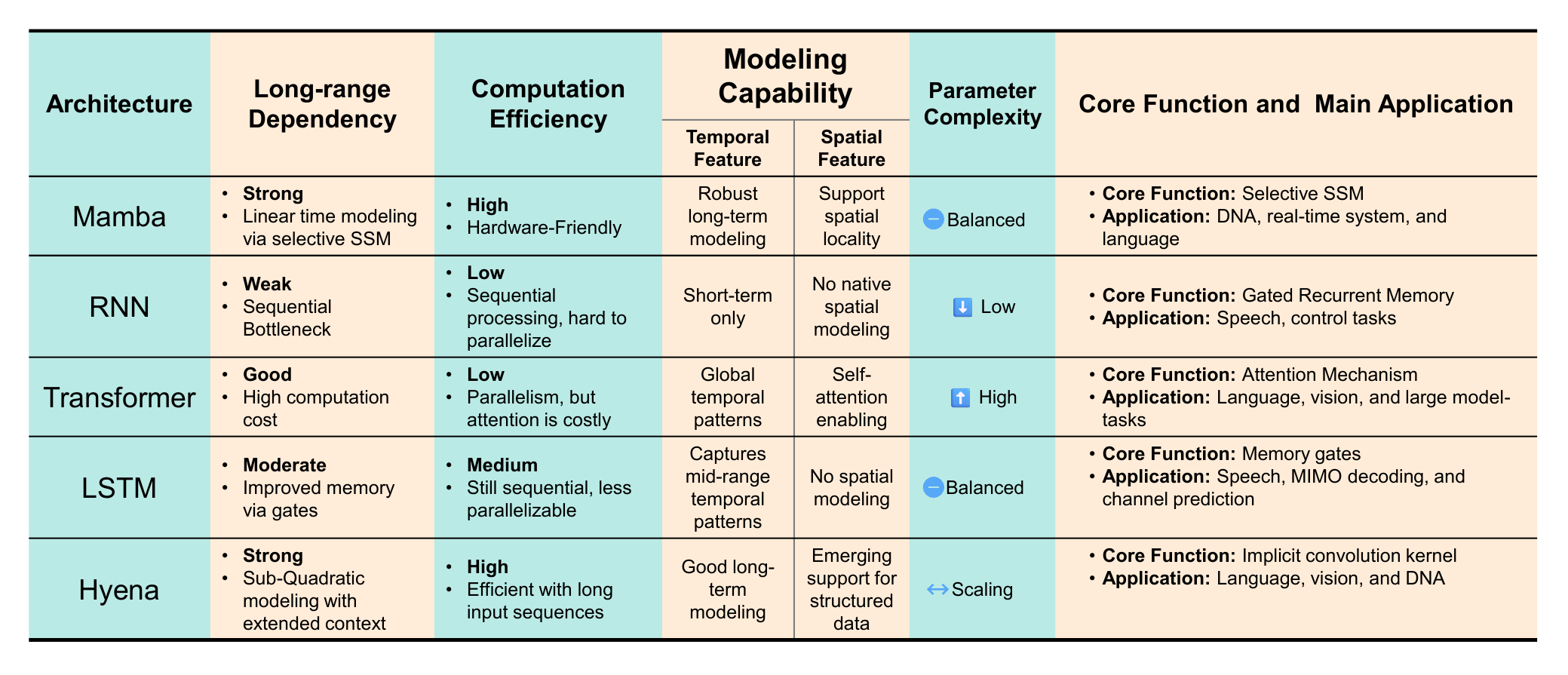}
\label{Table}
\end{table*}

\subsection{Overview of Mamba}
\begin{figure*}[t]
\begin{center}
{\includegraphics[ width=0.95\linewidth]{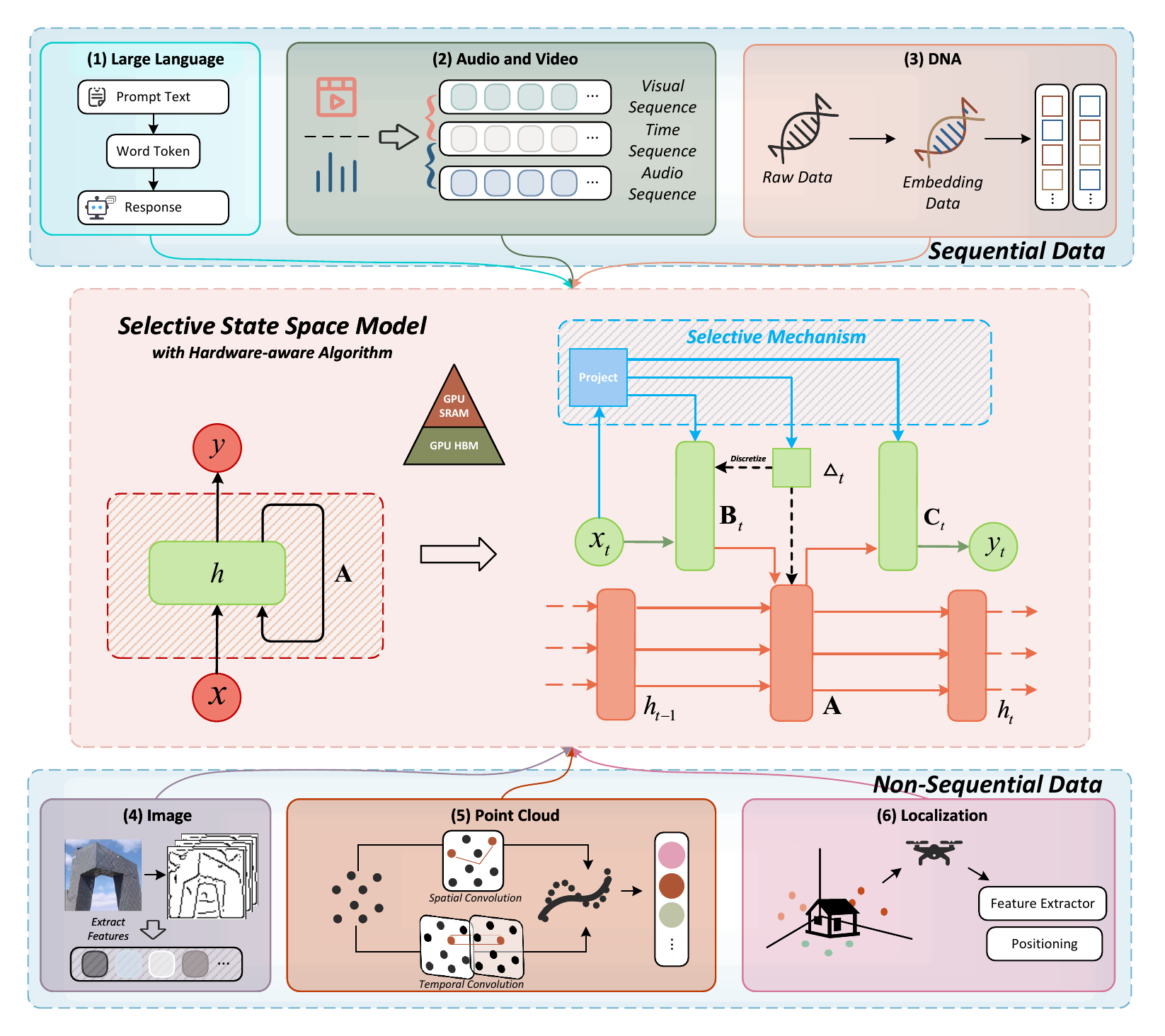}}
\caption{1) Illustration of the Mamba architecture and its core mechanism, i.e., Selective State Space Model. 2) Illustration of Mamba's working principles with representative applications in both sequential and non-sequential data domains.}
\label{SSM}
\end{center}
\end{figure*}

Mamba is based on the structured SSM, which is also improved by the incorporation of the selective state space\cite{Mamba} as illustrated in Figure \ref{SSM}. We first present a concise overview of SSM, and then emphasize the advantages of selective state space. 

SSM is a powerful framework widely used in time series modeling, signal processing and control systems. To  capture the dynamic evolution of the target system more effectively, the SSM  defines a latent representation for the system state, which encompasses the historical and current information of the system. Particularly, at the $t$-th time step, the system encodes its  input signal ${x_t}$ with the previous hidden state ${h_{t-1}}$ to generate an updated state representation ${h_t}$, which is further used to yield the desired observation $y_t$. Such a system state evolution is controlled by three key matrices: the state transition matrix ${\bf A}$ determines how ${h_{t-1}}$ contributes to ${h_t}$; the input influence matrix ${\bf B}_t$ modulates how external inputs (i.e., $x_t$) affect the system state; the observation matrix ${\bf C}_t$ maps the system state to its observation (i.e., $y_t$).


In fact, the system state acts as an internal memory mechanism, which enables the system to model temporal dependencies and predict future behavior. Despite its effectiveness in modeling sequential data and capturing temporal dependencies, the SSM faces long-range dependency and computational efficiency. Traditional SSM assumes the fixed state transition ${\bf A}$, which struggle to generalize in highly nonlinear and dynamic environments. Moreover, as the state updates over time, the repeated multiplications with the transition matrix may gradually induce the loss of critical information from distant past inputs, making it difficult to capture long-range dependencies effectively.

\begin{figure*}[t]
\begin{center}
{\includegraphics[ width=1\linewidth]{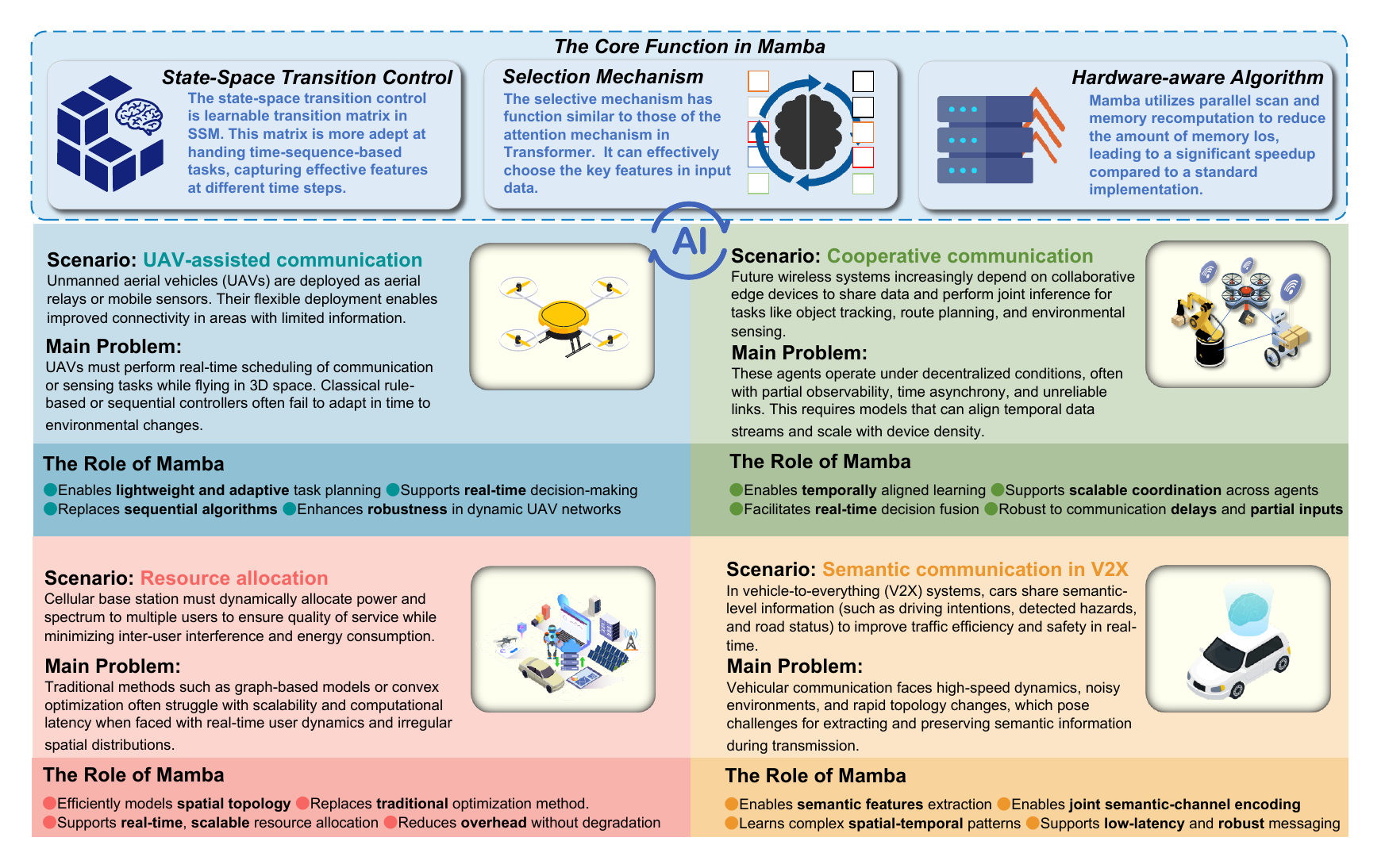}}
\caption{Overview of the core function for Mamba and its role to address the key challenges faced in some popular wireless communications (i.e., UAV-assisted communication, cooperative communication, resource allocation, and semantic communication).}
\label{Mamba-overview}
\end{center}
\end{figure*}

As illustrated in Figure. \ref{Mamba-overview}, Mamba augments the SSM with selective state space, which realizes flexible long-range dependency and thus, adapts to the input. To better understand the selective state space, we categorize its core functionalities into following three core aspects for a detailed discussion:
\subsubsection{\textbf{Dynamic State-Space Transition Control}} 


Mamba achieves the state evolution dynamically by introducing an input-dependent modulation. At different time steps, Mamba projects the input data as the update parameters of the state matrix, i.e., ${{\bf B}_t, {\bf C}_t}$, and ${\bf A}$. Besides, a time-step parameter ${\Delta_t}$ is also learnable based on the input data ${x_t}$. The system state is updated continuously with the input sequence for each transition being governed by the transition matrix. The matrix are learnable and dynamically adjusts throughout training and inference. This design ensures Mamba to adapt to diverse input conditions by the state-space transition control. 





\subsubsection{\textbf{Selection Mechanism}}
Inspired by the attention mechanism in Transformers, Mamba utilizes the selection mechanism to weighs its inputs at every steps. Then, these weighs modulate the impact of individual features on the state matrix update, effectively filtering and amplifying critical features while suppressing less relevant ones. This selective mechanism significantly enhances long-range dependency modeling and memory retention. Such selective processing has been proven to be essential in genomic analysis. For example, scMamba was proposed to process raw data by utilizing bidirectional Mamba blocks \cite{SCMamba}. It effectively learned the generalizable features of cells without relying on highly variable genes. 
\subsubsection{\textbf{Hardware-aware Algorithm}}
To tackle the inability to perform parallel computations in linear time invariance, Mamba utilizes three computation techniques, i.e, kernel fusion, parallel scan, and memory recomputation. Mamba fuses many operations, i.e., discretization, projection, and activation, into a single GPU high-bandwidth memory (HBM). It racially reduce the amount of memory IOs and leverages shared memory caching. Mamba transforms sequential state-space operation with a work-efficient parallel scan algorithm. 
To avoid some intermediate states reading between GPU memory cells in the forward pass, Mamba recalculates the intermediate states for backpropagation during long-sequence processing to reduce memory overhead. By efficiently processing high-dimensional audiovisual data, Mamba supports many real-time applications with low latency and high performance, including speech recognition, video understanding and dynamic scene analysis.

\subsection{Mamba's Applications}


Mamba's structured state-space architecture presents a unified modeling paradigm to capture both temporal dependencies in sequential data and spatial dependencies in non-sequential data. With its linear time complexity and continuous dynamics representation, Mamba is capable of adapting to various domains that require efficient context propagation, memory retention, and geometric awareness. We present two representative applications of Mamba in key domains, as follows.

\subsubsection{\textbf{Sequential Data}}
 Maintaining coherence over long text spans is crucial for generating meaningful output In Large Language models (LLMs) \cite{LLM}. Mamba enhances temporal consistency by replacing attention mechanism with selective SSM that integrates information over extended sequences more effectively. This approach alleviates the memory bottleneck inherent in Transformers while enabling longer context windows, which is essential for document-level understanding and code generation tasks. Moreover, for audio and video processing tasks \cite{MambaSurvey}, where signals exhibit strong coherence over time, Mamba's capability  to preserve long-range dependencies at scale enables precise modeling of rhythm, tone, and motion. Unlike recurrent models prone to vanishing gradients or self-attention models with high computational overhead, Mamba delivers  stable, low-latency temporal representations. This makes it well-suited for application such as real-time speech synthesis, music generation, and video captioning.

\subsubsection{\textbf{Non-Sequential Data}}
For specific  computer vision tasks \cite{CV}, Mamba captures hierarchical visual patterns with lightweight parameters. Its efficient long-range dependency modeling complements local convolutional operations, enhancing global contextual understanding with limited overhead. For 3D point cloud processing, where spatial irregularity and geometric continuity are critical, Mamba presents a promising alternative to graph-based or attention models. By encoding point-wise features into learnable latent states, Mamba enables the network to capture fine-grained spatial correlations and object structures without compromising  scalability. This is particularly valuable in autonomous driving, robotics and digital twin applications \cite{3D-point}. 

The above applications highlight the efficacy of Mamba’s core functionalities in extracting key features from structured data. Such a  powerful capability enables Mamba-based models to efficiently address spatial and temporal tasks. 

\subsection{Advantages of Mamba in Wireless Communications}

The wireless communication system embeds its spatial and temporal attribute into its task and data. Particularly, the radio frequency signals propagate in an open space and interact each other, and the signal processing is usually based on time-sequential data. Understanding and leveraging both spatial and temporal dependencies is crucial for address tasks from wireless networks. Compared with existing models, e.g., Transformer, Mamba offers some major advantages that make it suitable to wireless communication systems:
\subsubsection{\textbf{{Long-range Dependency Modeling}}}
Wireless communication systems involve plenty of estimation and prediction tasks, for example, CSI estimation, signal detection and radar sensing. The corresponding input data usually requires long-range temporal dependencies to reflect the continuous changes in wireless scenario (caused by signal fading, user mobility etc.) in order to achieve high precision. Mamba can address these tasks efficiently by establishing the relationship between the observed signals, and support indefinite long-range extrapolation over sequences exceeding one million tokens  \cite{Mamba}. Mamba enables the process of the (discretized) input data in two fashions, i.e., linear recurrence (where the inputs are seen one timestep at a time) and global convolution (where the whole input sequence is seen ahead of time) \cite{Mamba}. Note that the latter with more efficiency is unattainable with quadratic-complexity Transformers. Such an efficiency stems from Mamba's linear-time global convolution, enabling it more suitable for long sequences and real-time applications.

\subsubsection{\textbf{Spatial Feature extracting}}
The graph-topology wireless networks and the multi-functional RF signals complicate the spatial interaction among transmitters and receivers, such as inference, which results into multiple inter-node relation types. To capture and understand all relationships challenges the model architecture and training. Mamba can extract the spatial features through its selective state space, which dynamically adapts to heterogeneous node interactions in real time. Unlike GNNs that rely on predefined adjacency matrices or Transformer with fixed attention patterns, the input-dependent modulation allows it to learn implicit spatial dependencies conditioned on signal characteristics. Furthermore, its linear-complexity recursive processing enables efficient modeling of large-scale distributed antenna systems and reconfigurable intelligent surfaces (RISs) with thousands of spatially correlated elements. Mamba is critical for modern dense networks where traditional spatial models may fail to balance dynamic adaptability and computational cost.

\section{Application of Mamba in Wireless Systems}






This section explores the applications of Mamba in wireless systems. Particularly, we categorize the applications of Mamba in wireless systems into the following two categories:
\subsubsection{\textbf{Replacement of Traditional Algorithms}} Wireless signal processing tasks, including channel estimation, signal detection, and beamforming design, traditionally depend on iterative algorithms. However, they may encounter challenges in balancing trade-off between computational efficiency and effectiveness, especially in highly dynamic networks such as (unmanned aerial vehicle) UAV and (vehicle-to-everything) V2X scenarios. To  tackle this bottleneck, we propose customizing Mamba-based algorithms to replace their traditional counterparts. Particularly, the temporal and spatial features from the signals extracted by Mamba can be utilized for downstream signal processing tasks to enable high performance and rapid inference. Another key advantage is that Mamba  eliminates the need for  explicit mathematical formulation, and thus, enable data-driven processing of complex  tasks.

\subsubsection{\textbf{Enabler of Novel Paradigms}} Traditional communication systems employ  a hierarchical structure, where 
 functional modules (e.g., source coding and channel coding) operate in a separated manner. This structure enhances maintenance flexibility and reduces design complexity. However, the modular independence may lead to performance degradation compared to joint design approaches. To achieve integrated performance gain, we propose tailoring Mamba-based modules that integrate cascaded traditional modules. Particularly, Mamba can merge the multi-module functions in an end-to-end manner, effectively leveraging inter-module interactions. Additionally, Mamba-based modules enable adaptive information compression at diverse levels while preserving critical features.    

Subsequently, we present the Mamba-enabled designs for two typical tasks regarding to the two categories.

\subsection{Mamba-enabled Resource Allocation}

Wireless resource allocation tasks refer to assign limited resources (e.g., power budget and spectrum bandwidth) to enhance the system utilities (e.g., spectral  and energy efficiency) while satisfying the quality of service (QoS) requirements. These tasks are typically formulated as optimization problems. However, complex inter-element interactions and imperfect channel state information (CSI) pose significant challenges for developing efficient allocation schemes \cite{GNN}. Moreover, DL-based learning-to-optimize methods often require substantial training data, exacerbating challenges to training efficiency. 

For wireless networks, Mamba efficiently extracts inter-element spatial features to explore finer-grained trade-offs and inter-channel temporal features to predict or supplement the imperfect CSI. The information extracted by Mamba can be fed into downstream  DL layers for task-specific resource allocation tasks, enhancing learning efficiency and generalization performance. To develop Mamba-based resource allocation schemes, one can represent the task as a feature-to-decision mapping, and construct a neural architecture to realize the mapping with Mamba processes  input signals followed by decoders that correspond to the desired outputs. 



\subsection{Mamba-enabled Joint Source and Channel Decoding}


Semantic communication has emerged as a transformative paradigm in next-generation wireless networks, redefining transmission objectives from bit-level accuracy to semantic meaning preservation. As a core enabling paradigm, JSCD integrates  traditional modules to to achieve end-to-end optimization. At the transmitter, JSCD fuses the internal structure of source data (e.g., language redundancy in text or spatial correlation in images) with novel modeling paradigms of wireless channels (e.g., semantic-aware and data-driven frameworks) \cite{DeepSC}, while at the receiver, it directly decodes received signals into semantic meaning. Implementing JSCD requires a flexible modeling mechanism to simultaneously capture deep semantic dependencies in source content and characterize signal degradation caused by channel impairments.


 Mamba can be introduced into existing JSCD frameworks to enhance semantic modeling and robustness. By inserting Mamba layers into the semantic encoder and decoder, the model benefits from improved long-range dependency tracking and feature extraction. This integration allows the receiver to better infer semantic meaning from corrupted signals, with low overhead.

\section{Case Study}








This section presents  numerical results via case studies based on two Mamba-enabled designs. Particularly, we evaluate Mamba on two classic wireless tasks, i.e., wireless resource allocation and semantic communications. For the former, we integrate Mamba into the GNN-based model from \cite{GNN} to improve its inference efficiency in ultra-dense scenario; for the latter, we incorporate  Mamba to DeepSC from \cite{DeepSC} to enhance its performance  in the high-SNR region. 

\begin{figure*}[t]
\begin{center}
\begin{minipage}[c]{0.54\linewidth} 
    \centering
    \includegraphics[width=\linewidth]{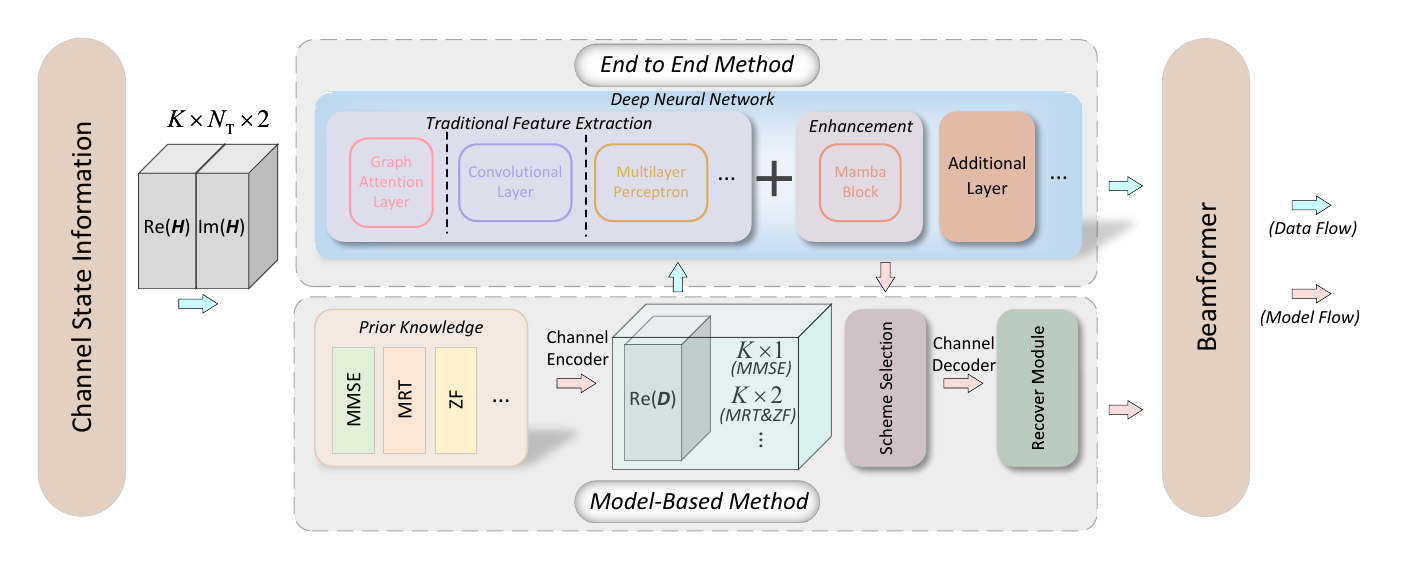}
    \label{fig:Resource}
\end{minipage}
\hfill
\begin{minipage}[c]{0.19\linewidth}
\vspace{0.6cm}
 \centering
    \includegraphics[width=\linewidth]{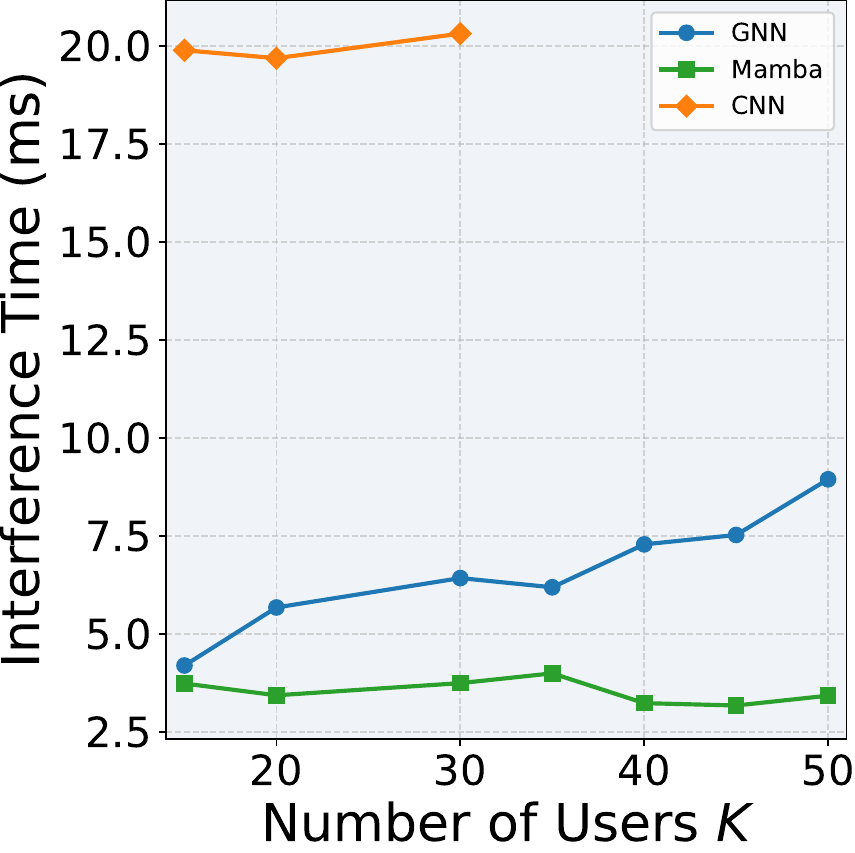}
    \label{fig:Time}
\end{minipage}
\hfill
\begin{minipage}[c]{0.25\linewidth}
    \rowcolors{2}{white}{gray!10}  
     \resizebox{\linewidth}{!}{
    \begin{tabular}{|c|c|c|}
        \hline
        \rowcolor{headerbg}        
        \textcolor{headertxt}{\textbf{Method}} &
        \textcolor{headertxt}{\textbf{GNN}} &
        \textcolor{headertxt}{\textbf{Mamba}}\\
        \hline
        $K$=15 & 86.25\% & 86.33\% \\
        \hline
        $K$=20 & 90.48\% & 90.54\% \\
        \hline
        $K$=30 & 98.42\% & 98.44\% \\
        \hline
        $K$=35 & 97.58\% & 97.54\%\\
        \hline
        $K$=40 & 93.61\% & 93.53\%\\
        \hline
        $K$=45 & 89.74\% & 89.58\%\\
        \hline
        $K$=50 & 86.05\% & 85.69\%\\
        \hline
    \end{tabular}}
\end{minipage}

\caption{Case study: Mamba for intelligent resource allocation with experimental setting following \cite{GNN}. Mamba is integrated into a GNN-based model to support both end-to-end and model-based resource allocation. With Mamba, the inference efficiency is enhanced, particularly for large-scale networks. 
}
\label{fig:combined}
\end{center}
\end{figure*}

Figure \ref{fig:combined} illustrates the detailed architecture of GNN-Mamba model for scalable energy efficient beamforming design for MISO networks. We adopt inference time and ratio to the optimal results as performance metrics. While GNNs with graph attention layers excel at capturing inter-user spatial correlations and enabling adaptive optimization under dynamic topologies, they face escalating computational complexity with network scales. Specifically, the GNN module suffers from quadratic attention complexity and limited scalability in ultra-dense scenarios. To address this, we propose the hybrid architecture by replacing part of the graph attention layers in the GNN model with Mamba blocks. Mamba is uniquely suited for scalable and efficient dependency modeling in large graphs, while maintaining constant inference time regardless of node count, ensuring adaptability to user scaling in practical networks. Experimental results demonstrate that the GNN-Mamba model significantly reduces inference time compared to baselines. On the other hand, the inference time of CNN and GNN exhibits monotonic increases  with user count, that of GNN-Mamba model remains near-constant. Regarding to energy efficiency, GNN and GNN-Mamba achieve close performance. These findings establish Mamba as a lightweight, scalable alternative for graph-based resource allocation in next-generation communication systems.

\begin{figure*}[t]
\begin{center}
\begin{minipage}[c]{0.74\linewidth} 
    \centering
    \includegraphics[width=\linewidth]{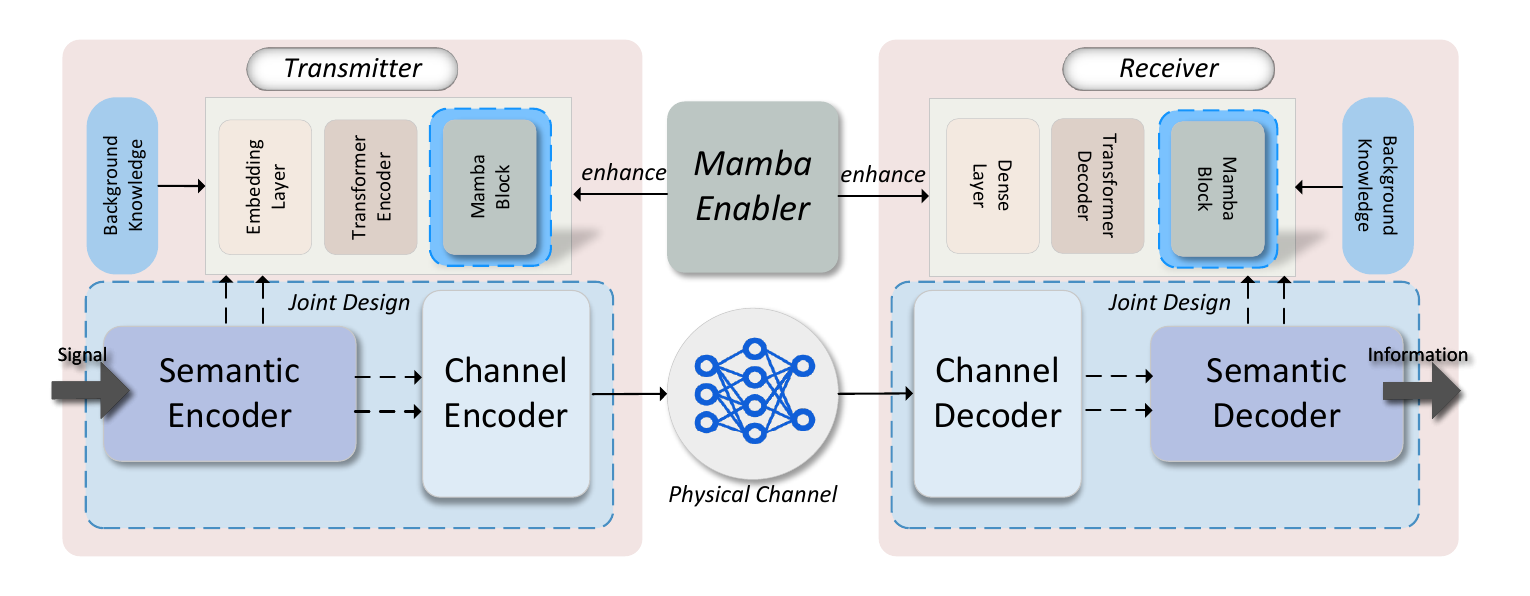}
    \label{fig:Semantic Communication}
\end{minipage}
\hfill
\begin{minipage}[c]{0.25\linewidth}
    \centering
    \rowcolors{2}{white}{gray!10}  
    \resizebox{\linewidth}{!}{
    \begin{tabular}{|c|c|c|}
        \hline
        \rowcolor{headerbg}        
        \textcolor{headertxt}{\textbf{Method}} &
        \textcolor{headertxt}{\textbf{DeepSC}} &
        \textcolor{headertxt}{\textbf{Mamba}}\\
        \hline
        SNR=0 & 52.13\% & 49.90\% \\
        \hline
        SNR=3 & 66.46\% & 64.07\% \\
        \hline
        SNR=6 & 80.16\% & 80.29\% \\
        \hline
        SNR=9 & 83.84\% & 87.44\%\\
        \hline
        SNR=12 & 87.54\% & 88.87\%\\
        \hline
        SNR=15 & 88.66\% & 90.98\%\\
        \hline
        SNR=18 & 90.63\% & 92.43\%\\
        \hline
    \end{tabular}}
\end{minipage}

\caption{Case study: Mamba for JSCD with experimental setting following \cite{DeepSC}. To enhance feature extraction, we embed Mamba into a Transformer-based JSCD architecture. }
\label{fig:combined2}
\end{center}
\end{figure*}

Figure \ref{fig:combined2} illustrates the the Mamba-enabled DeepSC to realize semantic communications. We adopt BLEU as the performance metric. DeepSC employs Transformer-based semantic and channel encoders for joint optimization of source and channel representations. However, Transformer architectures often suffer from high computational complexity, and their reliance on global attention mechanisms can be suboptimal when modeling long sequences with localized structures or hierarchical semantic dependencies.  To enhance the robustness and efficiency of semantic transmission, we propose inserting Mamba blocks as intermediate modules within both the semantic encoder and decoder frameworks. Specifically, following initial Transformer encoding, the semantic sequence is processed by a Mamba block before being fed into the channel encoder. On the receiver side, the reconstructed sequence from the channel decoder undergoes refinement through a Mamba layer prior to final semantic decoding. As observed, the Mamba-enabled DeepSC achieves performance gains in BLEU score across most SNR regions. This is because the sequential enhancement strategy allows Mamba to learn complementary temporal-spatial semantic features without altering the overall model architecture, effectively capturing long-range dependencies while maintaining computational efficiency.

It is noteworthy that the case studies have demonstrated Mamba's effectiveness as an augmenting module in existing models, yielding improvements in both feature enhancement and computational efficiency. These findings strongly indicate that Mamba can serve as a foundational building block for next-generation wireless intelligence.

\section{Future Directions}

The applications of Mamba in wireless communications and networking are still in their  early stages. The preceding analysis demonstrates Mamba's potential to enhance both feature representation and computational efficiency in future wireless communication systems. However, several critical challenges remain to be addressed, as detailed below.



\emph{1) Joint Task Scheduling in MEC}: In mobile edge computing (MEC) systems, task arrival patterns and network states exhibit dynamic temporal evolution. Mamba can model long-term workload fluctuations and user mobility to devise intelligent scheduling policies, thereby improving latency performance and resource utilization efficiency.

\emph{2) Interference Prediction in Heterogeneous Networks}: Ultra-dense network deployments face complex interference patterns. Mamba enables scalable and accurate modeling of temporal interference dynamics, supporting adaptive signal processing and frequency assignment strategies.

\emph{3) V2X Cooperative Perception and Fusion}: In V2X scenarios, sensing data from distributed nodes arrive with delays and noise. Mamba captures complex spatial and temporal dependencies among vehicles and high-precision maps to facilitate collaborative perception.

\emph{4) Collaborative UAV Communications}: For UAV swarms or aerial relaying systems, Mamba models trajectory correlations, task coordination mechanisms, and environmental feedback. This capability supports robust 3D coverage optimization and adaptive mission control.

\section{Conclusion}
This article has provided an overview of Mamba's applications in optimizing wireless networks, discussing its advantages and application frameworks. Specifically, we have analyzed Mamba’s suitability for wireless signal processing through its capabilities in long-range dependency modeling and spatial feature extraction. We have then introduced two core application frameworks, i.e., replacement of traditional algorithms and enabler of novel paradigms. Case studies have highlighted Mamba’s advantages as augmenting modules for intelligent resource allocation and semantic communications.

\bibliographystyle{IEEEtran}
\bibliography{IEEEabrv,ref}

\end{document}